\documentclass[twocolumn,showpacs,preprintnumbers,amsmath,amssymb,prl]{revtex4}
\usepackage{graphicx}
\usepackage{epsfig}
\usepackage{dcolumn}
\usepackage{bm}

\begin{document}

\title{Axionic Inflation from Large Volume Flux Compactifications}

\author{R.~Holman}
\email{rh4a@andrew.cmu.edu}
\author{Jimmy~A.~Hutasoit}
\email{jhutasoi@andrew.cmu.edu}
\affiliation{Dept. of Physics, Carnegie Mellon University,
Pittsburgh, PA 15213, USA.}

\date{\today}
\begin{abstract}
We find a general model of {\em single-field} inflation within the
context of type IIB string theory compactified on large volume
Calabi-Yau orientifolds with $h^{2,1}
> h^{1,1} = 2$. The inflaton is the axion part of the complexified
K\"{a}hler moduli and the resulting scalar power spectrum is red, and can easily be made
compatible with WMAP3 bounds on $n_s$. This model
overcomes the $\eta$-problem using gaugino condensates on wrapped
D7-brane while keeping the tuning of the parameters minimal.
\end{abstract}

\keywords{Inflation from string theory, large volume
compactifications in type IIB string theory}

\pacs{11.25.Mj,11.25.Wx,98.80.Cq,98.80.-k}

\maketitle

The recent advances in using both NS-NS and R-R fluxes to stabilize the
moduli of Type IIB Calabi-Yau (CY) compactifications \cite{gkp,kklt, grana} have sparked
intense interest in the possibility of generating the scalar potentials required for inflation (for previous attempts at string inflation, see \cite{stringinflation}). This has already resulted in some
interesting models, such as racetrack inflation \cite{racetrack},
which use a combination of flux generated potentials together with
non-perturbative potentials to construct a multi-field inflationary model.

Building on the work of Ref.\cite{kklt} (KKLT) and Balasubramanian
{\it et al.}\ \cite{vijay} (BBCQ), we show here that a {\em
single-field} inflationary model can be obtained from Type IIB
compactifications on CY orientifolds in the large volume limit. The
inflaton field is one of the axion fields arising as the real part
of a complexified K\"{a}hler modulus of the compactification. This
model solves the $\eta$-problem of supergravity inflation models in
a novel way: the smallness of $\eta$ comes from the size of a gauge
group $SU(N)$ on wrapped D7-brane that gives rise to gaugino
condensation.

The main idea is the following. BBCQ showed that in the large volume limit
and neglecting exponentially-suppressed contributions to the
potential, the potential has a non-supersymmetric minimum so that
one can stabilize the volume and some of the K\"{a}hler moduli. They
also argued that since this minimum is tachyon-free, one should be
able to fix all K\"{a}hler and complex structure moduli.

We proceed from where BBCQ left off and go back to the
exponentially-suppressed parts of the potential. This potential will
stabilize the remaining K\"{a}hler moduli and we will find that for
Calabi-Yau orientifolds with $h^{1,1} = 2$ (e.g.: ${\mathbb
P}^{4}_{[1,1,1,6,9]}$ \cite{denef}), this potential will give rise
to inflation.

This model can be thought of as a string version of natural
inflation models \cite{natural} that used Pseudo-Nambu-Goldstone
bosons as inflatons.

Let us first review the relevant aspects of moduli stabilization in
large volume compactifications. Type IIB string theory compactified
on Calabi-Yau orientifolds yields ${\cal N} = 1$ supergravity as the
four dimensional effective theory. Ignoring the gauge sectors, the
theory is specified by the K\"ahler potential and the
superpotential. Including the leading $\alpha'$ corrections
\cite{becker}, the K\"ahler potential is the form of
\begin{eqnarray}
K &=& - \log \left[-i \int_{M} \Omega \wedge \bar{\Omega} \right] -
\log \left[-i (\tau - \bar{\tau}) \right] \nonumber \\ & & -2 \log
\left[\, \frac{\xi}{2} \left( \frac{- i (\tau - \bar{\tau})}{2}
\right)^{3/2} + e^{-3 \phi_{0} /2} \, {\cal V} \, \right],
\end{eqnarray}
where $\tau$ is the axion-dilaton field, $\Omega$ is the (3,0)-form
of the Calabi-Yau, $\cal V $ is the classical volume of $M$ in units
of $l_{s} = (2\pi) \sqrt{\alpha'}$, and $\xi = - {\zeta(3)
\chi(M)}\slash ({2(2\pi)^{3}})$. We require that $\xi > 0$, or
$h^{2,1} > h^{1,1}$. Even though the superpotential $W$ \cite{GVW}
does not receive any $\alpha'$ corrections, it acquires a
non-perturbative part due to D3-brane instantons \cite{inst} or
gaugino condensation from wrapped D7-branes \cite{gaugino}:
\begin{equation}
 W = \int_{M} G_{3} \wedge \Omega + \sum_{i} A_{i} \, e^{ia_{i}\rho_{i}}.
\end{equation}
Here $G_{3} = F_{3} - \tau H_{3}$, with $F_{3}$ and $H_{3}$ are RR
and NS-NS 3-form fluxes, respectively, $A_{i}$'s are one-loop
determinants, $a_{i} = {2 \pi}\slash {M}$, with $M$ is a positive
integer. Also,  $\rho_{i} \equiv b_{i} + i \tau_{i}$ are the
complexified K\"{a}hler moduli consisting of the four-cycle moduli
$\tau_{i}$'s:
\begin{equation}
\tau_{i} = \partial_{t^{i}} {\cal V} = \frac{1}{2} D_{ijk} \, t^{j}
t^{k},
\end{equation}
and the axions $b_i$'s. The $t^{i}$'s measure the areas of
two-cycles, $D_{ijk}$ are the triple intersection numbers of the
divisor basis \cite{denef} and the classical volume is expressed as
\begin{equation}
{\cal V} = \frac{1}{6} D_{ijk} \, t^{i} t^{j} t^{k}.
\end{equation}

Since the non-perturbative effects in the superpotential are
exponentially small compared to the Gukov-Vafa-Witten part, we can
follow KKLT by approaching the problem of moduli stabilization in
two steps. First, we turn on the background fluxes to stabilize the
dilaton and complex structure moduli. Then, we incorporate the
non-perturbative effects and stabilize the K\"{a}hler moduli while
treating the dilaton and complex structure moduli as fixed.

The K\"{a}hler potential and the superpotential now become
\begin{eqnarray}
K &=& K_{cs} - 2 \log \left[\, \frac{\xi}{2} \left( \frac{- i (\tau
-\bar{\tau})}{2} \right)^{3/2} + e^{-3 \phi_{0} /2} \, {\cal V} \, \right], \nonumber \\
W &=& W_{0} + \sum_{i} A_{i} \, e^{ia_{i}\rho_{i}},
\end{eqnarray}
where all but the K\"{a}hler moduli  are taken to be fixed. From this we calculate the potential to be
\begin{eqnarray}
V &\equiv& V_{\alpha'} + V_{\rm np1} + V_{\rm np2},\nonumber \\
V_{\alpha'} &=& 3 \xi e^{K}  \frac{\xi^{2} + 7 \xi {\cal V} + {\cal V}^2}{({\cal V} - \xi) (2 {\cal V} + \xi)^{2}} \, |W|^{2}, \nonumber \\
V_{\rm np1} &=& e^{K} \, G^{\rho_{j} \bar{\rho_{k}}} \left(a_{j}
A_{j} a_{k} \bar{A_{k}} e^{i (a_{j} \rho_{j} - a_{k} \bar{\rho_{k}})} \right),\nonumber \\
V_{\rm np2} &=& e^{K}  i (a_{j} A_{j} e^{i a_{j} \rho_{j}} \bar{W}
\partial_{\bar{\rho_{k}}} K \nonumber \\ & & - a_{k} \bar{A_{k}} e^{-i a_{k} \bar{\rho_{k}}} W \partial_{\rho_{j}}K).
\end{eqnarray}

Let us consider the large volume limit where all $\tau_{i}
\rightarrow \infty$ except one, which we denote by $\tau_{s}$. One
of the conditions for $\tau_{s}$ is that this limit should be well
defined. In this limit, the potential reduces to
\begin{eqnarray}
V &=& V'_{\alpha'} + V'_{\rm np1} + V'_{\rm np2} + V_{\rm supp},\nonumber \\
V'_{\alpha'} &\sim& \frac{\xi}{{\cal V}^{3}} \, e^{K_{cs}}
|W_{0}|^{2},\nonumber \\
V'_{\rm np1} &=& e^{K} G^{\rho_{s} \bar{\rho_{s}}} a_{s}^{2} |A_{s}|^{2}
e^{-2 a_{s} \tau_{s}} \nonumber \\
&\sim& \frac{(- k_{ssj} t^{j}) a_{s}^{2} |A_{s}|^{2} e^{-2 a_{s}
\tau_{s}} e^{K_{cs}}}{{\cal V}}, \nonumber \\
V'_{\rm np2} &=& e^{K} G^{\rho_{s} \bar{\rho_{j}}} i a_{s} (A_{s}
e^{i a_s \rho_{s}} \bar{W} \partial_{\bar{\rho_{k}}} K \nonumber \\
& & - \bar{A_{s}} e^{- i a_s \bar{\rho_{s}}} W \partial_{\rho_{k}}
\bar{K}).
\end{eqnarray}
Here, $G_{\rho_s \bar{\rho}_j}$ is the K\"{a}hler metric on the size
moduli space and $V_{\rm supp}$ denotes the exponentially-suppressed
contributions to the potential. Since $V - V_{\rm supp}$ depends
only on $b_{s}$, $\tau_{s}$, and $\cal V$, we can take advantage of
the hierarchy in scales and approach the problem of stabilizing
K\"{a}hler moduli in two steps, just as in the KKLT construction.
Specifically, after fixing the dilaton and complex structure moduli,
instead of turning on all the non-perturbative effects, we first
only turn on $\tau_{s}$-dependent effects. Once $b_{s}$, $\tau_{s}$,
and $\cal V$ are stabilized, we treat them as fixed and go back to
$V_{\rm supp}$ to stabilize the rest of the K\"{a}hler moduli.

The only dependence on the axion $b_{s}$ is in $V'_{\rm np2}$.
Furthermore, extremizing the potential with respect to $b_{s}$ will
result in making $V'_{\rm np2}$ negative. BBCQ then showed that the
potential approaches zero from below for large $\cal V$ so that
there is a large-volume Anti de Sitter minimum. They also found that
at this minimum $\tau_s$ is stabilized at a small value, which goes
along with the earlier assumption of large volume limit.

This solution can be then uplifted to a de Sitter minimum by adding
anti-D3-branes \cite{kklt} or by using the supersymmetric D-terms
\cite{d-term}. The uplift potential is of the form
\begin{equation}
V_{\rm uplift} = \frac{\epsilon_{\rm uplift}}{{\cal V}^{3}},
\end{equation}
where $\epsilon_{\rm uplift} \geq 0$.

We are now ready to try to find the inflaton amongst all the moduli.
Consider Calabi-Yau orientifolds with two K\"{a}hler moduli $\rho_{s}$ and $\rho_{l}$.

After stabilizing $b_{s}$, $\tau_{s}$, and $\cal V$, the potential
reduces to
\begin{equation}
V = V_{0} + V_{\rm supp},
\end{equation}
where $V_0$ is a constant. Since ${\cal V} = {\cal
V}(\tau_{s},\tau_{l})$, this means that $\tau_{l}$ has also been
stabilized and so $V_{\rm supp} = V_{\rm supp}(b_{l})$ and
\begin{eqnarray}
&V_{\rm supp}& = \frac{e^{K} G^{\rho_{l} \bar{\rho_{j}}} a_{l} t_{j}
e^{- a_{l} \tau_{l}}}{2 {\cal V}} \left(A_{l} \bar{W_{0}} e^{i a_{l}
b_{l}} + \bar{A_{l}} W_{0} e^{- i a_{l} b_{l}} \right) \nonumber  \\
&=&  \frac{e^{K} G^{\rho_{l} \bar{\rho_{j}}} a_{l} t_{j} e^{- a_{l}
\tau_{l}}}{ {\cal V}} |A_{l} W_{0}| \cos(\theta_{W}
- a_{l} b_{l}),
\end{eqnarray}
where $\bar{A_{l}} W_{0} \equiv |A_{l} W_{0}| e^{i \theta_{W}}$.

The range of $V_{0}$ is model-dependent. However, for $\epsilon_{\rm
uplift} = 0$, $V_{0} < 0$. Furthermore,  for some values
$\epsilon_{\rm uplift} > 0$, $V_{0} > 0$, though in this case,
$\epsilon_{\rm uplift}$ must be sufficiently small such that the
minimum is not entirely wiped out. Since the value of $V_{0}$ is a
continuous function of $\epsilon_{\rm uplift}$, one should be able
to tune the parameters to get $V_{0}$ to be exactly zero regardless
of the geometry.  Assuming that this has been done, we then proceed
by setting $\theta_{W} = 0$ and get $V = ({\rm constant}) \, \cos(a_{l} b_{l})$.

As is typical in supergravity theories, the kinetic term for the
axion $b_{l}$ is not canonical. However, since we have fixed the
other moduli, we can rescale $b_{l} \rightarrow \lambda b_{l}$ with
$\lambda = \sqrt{2 / G_{\rho_{l} \bar{\rho_{l}}}}$, which is constant, to
make the kinetic term canonical. The equation of motion in an FRW background is:
\begin{equation}
\ddot{b}_{l} + 3 H \dot{b}_{l} + \frac{\partial V}{\partial b_{l}} =
0 ; \ \ \ \  3 H^{2} = \frac{1}{2} \dot{b}_{l}^{2} + V(b_l),
\end{equation}
where dots denote cosmic time derivative and in our units, $8\pi G_N=1$.

The slow-roll parameters are given by \cite{riotto}
\begin{eqnarray}
\epsilon &=& \frac{1}{2} \left( \frac{\partial V\slash \partial b_{l}}{V}
\right)^{2} = \frac{a_{l}^{2}}{2 \lambda^{2}} \tan^{2}\left(\frac{a_{l}
b_{l}}{\lambda}\right),\nonumber \\
\eta &=& \frac{\partial V^{2} / \partial^{2} b_{l}}{V} = -
\frac{a_{l}^{2}}{\lambda^{2}}.
\end{eqnarray}
Notice that $\eta$ is constant, which means that once we get a small
$\eta$, the validity of slow-roll is solely determined by
$\epsilon$. Of course, this is not true for $V_{0} \ne 0$.

We can rewrite the equations of motion in terms of the number of
$e$-folds $N$, where $N = \int H dt$:
\begin{equation}
 H^{2}  = \frac{V}{3 - \frac{1}{2} {b^{\prime}_{l}}^{2}}, \nonumber
 \end{equation}
 \begin{equation}
 b^{\prime \prime}_{l} + \left(3 - \frac{1}{2} {b^{\prime}}_{l}^{2} \right) b^{\prime}_{l}
 -
\frac{a_{l}}{\lambda} \left(3 - \frac{1}{2} {b^{\prime}}_{l}^{2}
\right) \tan \left(\frac{a_{l} b_{l}}{\lambda} \right) = 0.
\end{equation}

Using $\eta = -{a_{l}^{2}}\slash{\lambda^{2}}$, we can rewrite
$\epsilon$ as a function of $\eta$
\begin{equation}
\epsilon = \frac{1}{2} \left |\eta  \right | \tan^{2}
\left(\sqrt{\left | \eta \right |} \, b_{l} \right).
\end{equation}
To get inflation, we need to tune $\eta$ to be small and let the
axion $b_{l}$ roll from near the top of the potential. Now $\lambda$
depends on the geometry. However, it is constant and since $a = 2
\pi\slash M$, where $M\in {\mathbb Z}^{+}$ can be a large integer,
tuning $\eta$ is not a serious problem. It is interesting that our
ability to get a small $\eta$ comes mainly from the {\em discrete}
parametrization of the gauge field $SU(M)$ on the wrapped D7-brane.

\begin{figure}[t]
\raggedleft \centerline{ \epsfxsize=3.5in \epsfbox{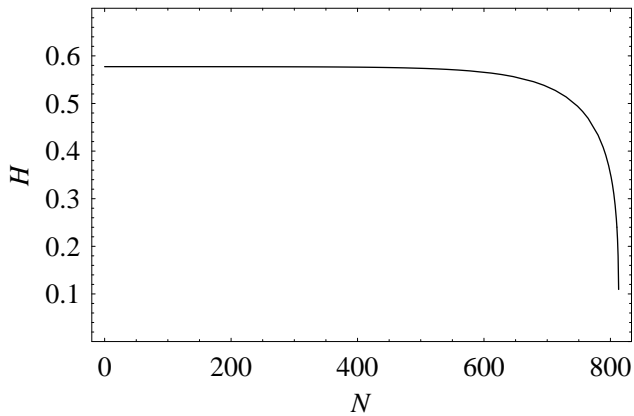}}
\caption{$H = \frac{\dot{a}}{a}$, where $a$ is the scale factor, is
constant. Thus the scale factor is exponentially increasing.}
\label{fig:hubblepar}
\end{figure}

Furthermore, we have a constraint from the observation of the scalar
power spectrum, namely that the spectral index $n_s = 0.951^ {+
0.015}_{-0.019}$ \cite{wmap}. For slow-roll inflation, we have the
constraint \cite{riotto}
\begin{equation}
\label{eq:eta}
n - 1 = 2 \eta - 6 \epsilon = -\left | \eta \right | \left(2+3\tan^2\left(\sqrt{\left |\eta\right |} b_l\right)\right)<0.
\end{equation}
This tells us that we need $\left | \eta \right | \lesssim 0.03$.
For concreteness, $\eta = - 0.006$. If we take as initial conditions
$b_{l}(N = 0) = 0.1$ and $b'_{l}(N = 0) = 0$, we get $800$ e-folds
or so of inflation (Fig.\ref{fig:hubblepar}).

\begin{figure}[t]
\raggedleft \centerline{ \epsfxsize=3.5in \epsfbox{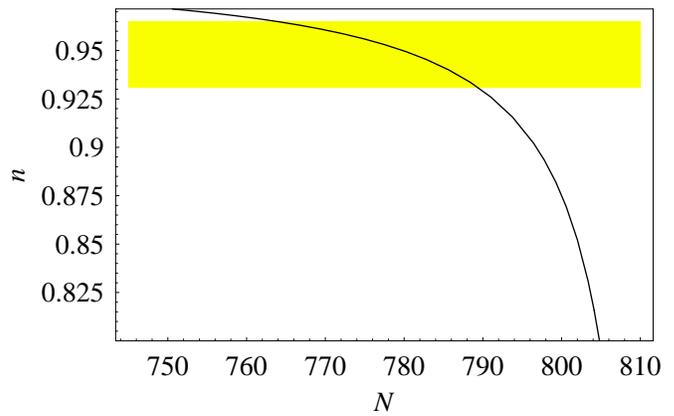}}
\caption{Scalar spectral index from axionic inflation for the last
$60$ e-folds before the end of inflation. The shaded region
corresponds to the bounds from  WMAP3.} \label{fig:spectralindex}
\end{figure}

The spectral index for the last 60 $e$-folds before the end of
inflation is plotted in Fig.\ref{fig:spectralindex}, and we can see
that the observational constraint is compatible with our results.

Since $V_{\rm supp}$ is exponentially suppressed, the scale of inflation
is far below the $3 \times 10^{16}$ GeV threshold required to
produce observable tensor contribution to the power spectrum.

We can get sufficient inflation in this model for a large range of
initial values of the axion field $b_l$, although they have to be
relatively near the top of the potential. How likely this is to
happen requires further analysis.

String theory gives us a large number of scalar fields, one or more
of which might have been the driving force of inflation in the early
universe. However, it is rare to find a single field inflation model
from string theory as we have done in this paper. The reason is that
in general, at every stage of moduli stabilization, we are dealing
with more than one scalar field. At each stage, it is tempting to
constrain all fields, except one or two, along the minimum and use
the rest, let us call them dynamical field(s), as inflaton(s). For
example, when fixing $b_{s}$, $\tau_{s}$, and $\cal V$, one might
want to constrain $b_{s}$ and $\cal V$, and only treat $\tau_{s}$ as
a dynamical field. However, in general, this is not valid because:
\begin{enumerate}
\item  Since we have not finished stabilizing all fields in that stage,
the constrained fields are still functions of the dynamical fields.
For example, before stabilizing $\tau_{s}$, extremizing the
potential with respect to $b_s$ and $\cal V$ only resulted in
expressing them as functions of $\tau_{s}$, but not as fixed values.
Thus, as the dynamical fields evolve, the constrained fields evolve
too. Therefore, since all fields are coupled to gravity, in general,
we can not neglect the dynamics of the constrained fields in doing
the analysis for inflation.

\item  We cannot assume {\it a priori} that as the dynamical fields evolve,
the constrained fields will remain constrained. It is
more likely that these fields will oscillate around their
respective constrained forms. Whether the oscillation can be
neglected requires careful analysis of the equations of motion for
the constrained fields.
\end{enumerate}

What is different in our situation is that we have used the fact
that $V_{\rm supp}$ is exponentially suppressed to approach the
problem of moduli stabilization in three stages, where in the last
stage we only have to deal with the axion $b_{l}$, which we then use
as the inflaton.

In order to do a geometry-independent argument, we have tuned
$V_{0}$ to be zero. As we have discussed above, this is not an
unnatural thing to do from the perspective of flux
compactifications. Furthermore, $V_{0} = 0$ is not a necessary
condition for inflation either. We can still produce inflation with
$V_{0} < 0$ as long as the top of potential is still above zero.
Obviously, $V_{0} > 0$ will not prevent inflation from happening.
However, in order for inflation to stop, $V_{0}$ should be
sufficiently small such that the bottom of the potential is not
above zero.

The theory of reheating is still an open problem in this model.
Since we are not working on a specific geometry, we are not going to
discuss the problem of reheating in this paper. However, with the
recent progress in the study of phenomenological implications in
particle physics that emerge from large volume compactifications,
esp. concerning mechanism of supersymmetry breaking \cite{kerim}, it
would be interesting to work on a specific model that has both
inflationary sector and SUSY breaking sector, and address the issue
of reheating. We hope to do so in a future publication.

\section{Acknowledgement}
We would like to thank C.~P. Burgess for helpful discussions. This
work was supported in part by DOE grant DE-FG03-91-ER40682.

\end{document}